\documentclass[final,preprint,aps,showpacs,fleqn,nofootinbib]{revtex4-1}
\usepackage[toc]{appendix}
\usepackage{epsfig,graphicx}
\graphicspath{./}
\usepackage{amsmath}
\usepackage{hyperref}
\hypersetup{colorlinks,linkcolor=red, citecolor=green}

\newcommand{\be}{\begin{equation}}
\newcommand{\ee}{\end{equation}}
\newcommand{\bea}{\begin{eqnarray}}
\newcommand{\eea}{\end{eqnarray}}
\newcommand{\ba}{\begin{array}}
\newcommand{\ea}{\end{array}}
\newcommand{\tr}{{\rm tr}}
\newcommand{\Tr}{{\rm Tr}}

\usepackage[normalem]{ulem} 
\usepackage[dvipsnames]{xcolor} 
\renewcommand\sout{\bgroup \color{red} \ULdepth=-.5ex \ULset}

\usepackage[left=3cm,top=2cm,bottom=2cm,right=2cm]{geometry} 
\usepackage[compat=1.0.0]{tikz-feynman}
\usetikzlibrary{shapes,arrows,positioning,automata,backgrounds,calc,er,patterns}
\tikzfeynmanset{every blob={/tikz/fill=none, /tikz/inner sep=2pt},every vertex={dot,minimum size=2pt}}

\begin{document}

\title[Gluons, Light and Heavy Quarks in the Instanton Vacuum]{Gluons, Light and Heavy Quarks in the Instanton Vacuum}

\author{M.~Musakhanov$^1$, U.~Yakhshiev$^{1,2}$}
\address{${}^1$Theoretical Physics Department, National University of Uzbekistan, Tashkent 100174, 
Uzbekistan\\
${}^2$Department of Physics, Inha University, Incheon 22212, Korea}
\email{musakhanov@gmail.com}

\date\today

\begin{abstract}
In this review we will concentrate on the nonperturbative effects 
in the properties of hadrons made from the light-heavy and 
heavy-heavy quarks in the framework of Instanton 
Liquid Model (ILM)  of  QCD vacuum.
We briefly discuss the main features of ILM and its applicability
in the heavy quark sector. The properties
of gluonic systems, light and heavy quark correlators 
in the instanton background are also analyzed.  
Consideration of the both, perturbative and 
nonperturbative, gluon effects in the instanton background 
for the single heavy quark will lead to the mass shift due to the 
direct-instanton nonperturbative 
and ILM modified perturbative contributions, respectively.
For the interacting  heavy quark-antiquarks, the potential
consists the direct instanton induced part and the 
one-gluon exchange (OGE) perturbative part. 
OGE interactions are screened 
at large distances due to the nonperturbative dynamics. 
We discuss the estimations of instanton contributions in the 
phenomenological Cornell type potential model.  As related to the experimental
data we discuss the charmonium properties 
and the role of instanton effects in their spectra and
transitions.
We discuss also the main features of heavy-light quarks systems 
in the ILM. As an example, it is considering the process of pions emission by
exited heavy quarkonium states. 

\end{abstract}


\maketitle
\thispagestyle{empty}

\section{Introduction}

With upgrading the Large Hadron Collider's (LHC)  accelerating and 
detecting facilities in the current decade it will be available more and 
more data coming from the experiments having a reach 
information on the properties of 
hadrons\,\cite{Apollinari:2017cqg}.  Particular 
interest during these experiments is focused on the properties of heavy hadrons containing 
the heavy quarks inside while they naturally allow to probe 
all energy regimes of Quantum Chromodynamics (QCD) due to the 
existance of heavy particle states and their decay
modes to another more lighter particle states~\cite{Choi:2003ue,
Aubert:2004ns, Aubert:2005rm, Abe:2007jna, 
  Choi:2007wga, Belle:2011aa, Liu:2013dau, Ablikim:2013mio,
  Ablikim:2013wzq, Aaij:2013zoa,Ablikim:2013xfr,Aaij:2014jqa, 
  Aaij:2015zxa,Yuan:2015kya,Brambilla:2010cs}. 
The weak decays of heavy hadrons give the useful information 
in verification of the Standard Model and probing the physics 
beyond. The heavy hadron physics also gives important information 
about the deconfinement region (e.g. formation of quark-gluon plasma) 
and that part of strong interaction forces which remains still
not well understood as related to the hadron properties 
in confinement region.

From the point of view of their structures one can classify the heavy 
hadrons into two classes, e.g. open heavy flavor (with one or two 
heavy quarks inside) and hidden heavy 
flavor (with one heavy quark and the corresponding heavy 
antiquark inside) systems. 
In both cases a nonrelativistic approximation 
to the heavy quark properties seems to be reasonable. 
This is due to the fact that in  a reduced system the heavy quarks 
can be considered as a nonrelativistic constituents and the 
relativistic corrections could be taken into account
by developing some systematic approach in terms of the 
appropriate parameters and available constants~(e.g. see Ref.\,\cite{Neubert:1993mb}).

In this context, it is also necessary to note that the 
properties of open heavy flavor hadrons seem to be 
mainly governed by the light quark properties in the system. In a more fundamental approach to the properties of such hadrons
the phenomenon of spontaneous breaking of the chiral symmetry
should be taken into account in an appropriate way. 
This is due to the fact that the
properties of light quarks mainly governed by this phenomenon.
In such a way, they are related to the structure of QCD vacuum and 
the near vacuum phenomena. Consequently, the trace of  
the nonperturbative region may be essential in the open heavy
flavor systems not only for their decay modes into the lighter 
hadrons but also in their static properties. In addition to this
and in a more fundamental level
the properties of light quarks in the heavy 
hadron sector should be considered in the framework 
of relativistic theory. Therefore, an applicability of
nonrelativistic potential approaches to the properties of open heavy 
flavor hadrons and to the decay processes of all heavy hadrons 
seems to be not well justified.

In contrast, the properties of hidden heavy flavor systems could be
accounted in the nonrelativistic approximation.
In particular, the heavy quarkonium (a system with one heavy 
quark and one heavy anti-quark) plays an 
essential role during these studies.  Nevertheless, an interaction 
forces in the heavy quark-antiquark systems may still have 
response from the nonperturbative region. 
The proper account of
nonperturbative effects in the potential models to the charmonium 
properties may shed light into the origin of parameters of the 
model and may improve the theoretical calculations not only 
qualitatively but also quantitatively. The corresponding discussions
are our aim in the 
present review and we will perform our task on a basis 
several works\,\cite{Diakonov:1989un,Turimov:2016adx,Yakhshiev:2018juj,Musakhanov:2020hvk} developed in the framework of 
instanton liduid model (ILM) of QCD vacuum.

We perform our review in the following way. 
In the next Section~\ref{sec:ILM}, we briefly discuss the main 
futures of ILM in relation to the phenomenological observations
and the applicability of model in the heavy quark sector.
In Section~\ref{sec:QuarkCor} we discuss the heavy quark correlators
in the instanton medium by taking into account also the
perturbative corrections. After, we briefly discuss a gluon 
propagation in instanton vacuum in Section~\ref{sec:ginILM}.
We analyze a heavy quark propagator in instanton medium
 in  Section~\ref{sec:HQP} and discuss the corresponding contributions 
 to the heavy quark mass. In Section~\ref{sec:PotMod},
 we briefly mention about the potential approaches to the 
 quarkonium properties. The contribution to the heavy-quark
due to instanton effects are discussed in Section~\ref{sec:IconPot}
and order of these effects are discussed in Section~\ref{sec:InsEffects}.
In Section~\ref{sec:hlqinILM} we discuss the main features of
the heavy-light systems in the instanton vacuum and crudely
estimate couplings for the pion transitions in charmonia states and get
sizable corrections ( $\sim 20\%$) to the
dipole approximation for the process $\psi(2S)\rightarrow J/\psi\, \pi^+\pi^-$ . 
Finally, in Section~\ref{sec:SumOut} we summarize our discussions.

\section{Model and its parameters}
\label{sec:ILM}

\subsection{ILM parameters}

QCD vacuum is quite nontrivial non-perturbative vacuum state
characterized by the nonvanishing gluon and quark condensates (e.g.
see Ref.\,\cite{Ioffe:2005ym}). There are different models of
QCD vacuum and the instanton liquid model is one such 
models (e.g. see
Ref.\,\cite{Diakonov:2002fq}). ILM nicely 
describes the spontaneous breakdown of chiral symmetry which is one
of the essential features of strong interactions in the 
nonperturbative region. Instanton model is also related to the rich 
topological structures of QCD vacuum and, although many scientists
skeptic about that, may still be relevant to the confinement if not 
directly in relation with other mechanisms. 

An instanton is  a classical solution of Yang-Mills equations in 
the 4-dimensional Euclidean space. The potential part of the Yang-Mills
action has the periodic structure in the functional space along the 
collective coordinate direction which is called the Chern-Simons (CS) 
coordinate and the minimum energy state is infinitely
degenerated in that direction. Therefore, QCD vacuum can be 
considered as the lowest  energy quantum state of the one-dimensional periodic
crystal along the CS coordinate~\cite{FJR1976,Jackiw:1976pf}.
The instanton is a tunneling mechanism in one direction (let to say 
in forward direction) between the different Chern-Simons states
corresponding to the degenerate vacuum while the anti-instanton is 
transition in opposite direction~\cite{Belavin:1975fg}.

The instanton is described by its collective
coordinates denoted as $\xi_{I}$: the position in 
4-dimensional Euclidean 
space $z_{I}$, the instanton size $\rho_{I}$ and the 
SU$(N_{c})$ color orientation 
given by the unitary matrix $U_{I}$,
$4N_{c}$ variables altogether.\footnote{Hereafter, we drop 
the subscript $I$ for the convenience
and note that $N_c$ is the number of 
colors.}  The anti-instantons are also characterized by the similar 
coordinates. There are two main parameters in ILM 
-- the average instanton size
$\bar\rho$ and the inter-instanton distance $R$. The latter one  
describes the density of instanton media $N/V\equiv
1/R^4$\footnote{Here $N$ is the 
total number of instantons.} and phenomenologically  related 
to the gluon condensate~\cite{Shifman:1978bx} 
\begin{equation}
\frac{N}{V}\simeq\frac{1}{32\pi^2}\langle F_{\mu\nu}^aF_{\mu\nu}^a\rangle=
\frac{1}{16\pi^2}\langle {\bf B}^2-{\bf E}^2\rangle\simeq (200\,{\rm MeV})^4.
\end{equation}
Consequently, one has $R\simeq 1\,{\rm fm}$.

\begin{figure}[hbt]
\begin{center}
\includegraphics[scale=0.75]{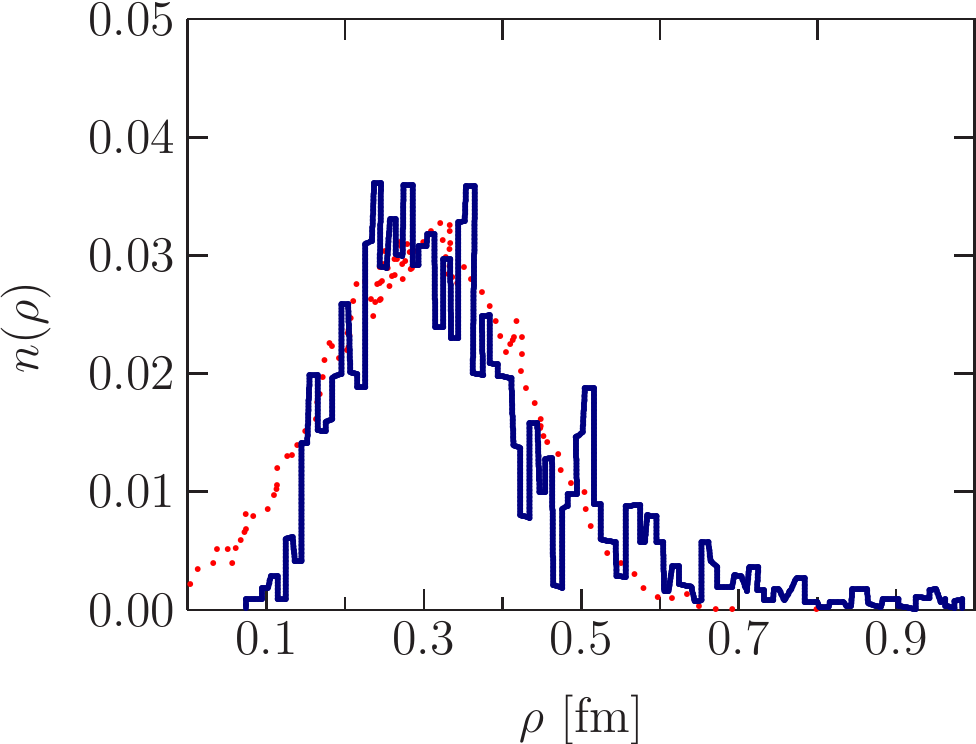}
\end{center} 
\caption{The instanton size distribution 
function $n$ depending on the instanton size parameter $\rho$.
The dots correspond to the calculations in the framework of
ILM while the continuous lines
correspond to the lattice simulations~\cite{Millo:2011zn}.}
\label{instantonsize} 
\end{figure}

From other side, 
the instanton size distribution $n(\rho)$
has also been studied by lattice simulations~\cite{Millo:2011zn}. It
 is shown in Fig.\ref{instantonsize}
where the calculations
in framework of ILM model are also given 
for the comparison. One can see that, at 
the relatively large values of parameter $\rho\sim R$ providing
more intensive overlapping of instantons,  the distribution function 
$n(\rho)$ is suppressed.  Rather narrow distribution is localized 
around $\rho\simeq0.35$\,fm corresponding to the average size
$\bar\rho$.  Therefore, in practical calculations one can replace 
all instantons by the average-size instanton. This also provides 
a simple sum-ansatz for the total instanton field $A(\xi)=\sum_iA_i(\xi_i)$
expressed in terms of the single instanton
solutions  $A_i(\xi_i)$ while $\bar\rho$ becomes much
smaller than $R$, $\bar\rho/R\simeq 1/3$. Although, nothing can 
prevent some instantons to be large in size and,  
in such a way,  may lead to the overlapping 
them with each other. However, the 
phenomenological estimates shows that the majority of instantons
are well-isolated
\begin{equation}
\bar\rho\simeq\frac{1}{3}\,{\rm fm},\qquad R\simeq1\,{\rm fm}.
\label{eq:SetInstanton}
\end{equation}
These values were  confirmed by the theoretical variational 
calculations~\cite{Diakonov:2002fq,Schafer:1996wv,shuryak2018}
and the lattice simulations of the QCD vacuum~\cite{Chu:1994vi,Negele:1998ev,DeGrand:2001tm,Faccioli:2003qz}.

\subsection{Heavy hadrons' core sizes}

In order to elaborate the instanton liduid model, which was a 
powerful tool
in the light quark sector, to the heavy-light and 
heavy-heavy quark systems one should analyze the applicability 
range of model parameters in comparison with 
the hadron's quark core sizes.

Concentrating back again to the instanton size distribution function 
$n(\rho)$ shown in Fig.\,\ref{instantonsize}, one can note 
that the large-size tail  
becomes important in the confinement regime of 
QCD. Here in order to take instanton phenomena more accurately,
one should  replace Belavin-Polyakov-Schwarz-Tyupkin instantons 
by Kraan-vanBaal-Lee-Lu 
instantons~\cite{Kraan:1998kp,Kraan:1998pm,Lee:1998bb}
 described in terms of  dyons. 
In such a way, one gets a natural extension of instanton liquid 
model, i.e. liquid dyon model (LDM)\,\cite{Diakonov:2009jq,Liu:2015ufa,Liu:2015jsa}. The extended model will allow to reproduce 
confinement{\textendash}deconfinement phases. 
The small size instantons can still be described in terms of their collective coordinates. For comparison, 
the average size of instantons in liduid dyon model is 
$\bar{\rho}\approx0.5\,{\rm fm}$
\cite{Diakonov:2009jq,Liu:2015ufa,Liu:2015jsa}, while in 
instanton liquid model the average size is 
$\bar{\rho}\approx0.3\,{\rm fm}$ as already discussed above.
In actual calculations, one can neglect the effect of 
size distribution's width 
and for a simplicity consider the instanton size $\rho$
always equal to its average value, $\rho=\bar\rho$.
Hereafter, we also always use the average value of instantons 
in our calculation.

At the typical values of the ILM parameters
given in Eq.\,(\ref{eq:SetInstanton}), one can estimate
the QCD vacuum energy density which takes the nonzero 
value, $\epsilon\approx-500\,
{\rm MeV/fm^3}$\,\cite{Schafer:1996wv,shuryak2018}.
Due to the instanton fluctuations it occurs a spontaneous breakdown 
of chiral symmetry
which plays the pivotal and significant role in describing the 
lightest hadrons and their interactions. 
In such a way, ILM succeeded to reproduce spontaneous 
symmetry breaking  and explain the corresponding hadron physics 
at the light quark sector. For more details, see  
reviews~\cite{Diakonov:2002fq,Schafer:1996wv,shuryak2018}
and for some other applications Refs.~\cite{Goeke:2007bj,Goeke:2007nc,Goeke:2010hm,Musakhanov:2012zm,Musakhanov:2018sdu}.

In order to understand the applicability of instanton liquid model 
in the heavy quark sector, one should compare 
the typical sizes of quarkonia and ILM model parameters. 
For example, the sizes of heavy quarkonia are relatively small~\cite{Digal:2005ht,Eichten:1979ms} (see Table~\ref{Quarkoniumsizes}). One can see, that this is more 
 pronounced in the case of
low laying states $r_{J/\psi}$ and $r_{\Upsilon}$. 

\begin{table}[h]
\caption{Masses and  sizes of quarkonium
in the non-relativistic potential model
\cite{Digal:2005ht}.}
\footnotesize
\begin{center}
\begin{tabular}{@{}c|ccc|ccccc}
\hline 
Characteristics of states
&\multicolumn{3}{c|}{Charmonia}
&\multicolumn{5}{c}{Bottomonia}\\
& $J/\psi$  & $\chi_{c}$  & $\psi'$  & $\Upsilon$  & $\chi_{b}$ & $\Upsilon'$ & $\chi_{b}'$ & $\Upsilon^{''}$ \tabularnewline
\hline 
mass {[}GeV{]}  & 3.07  & 3.53  & 3.68  & 9.46  & 9.99 & 10.02 & 10.26 & 10.36 \tabularnewline
size $r$ {[}fm{]}  & 0.25  & 0.36  & 0.45  & 0.14  & 0.22 & 0.28 & 0.34 & 0.39 \tabularnewline
\hline 
\end{tabular}
\end{center}
\label{Quarkoniumsizes}
\end{table}
Estimations of nucleon's quark core sizes also give 
the similar results $r_{N}\sim0.3-0.5$\,fm~\cite{He:1986yq,Weise,Tegen}. 
While the quark core of hadrons
are relatively small, 
one may conclude that the core parts of hadrons are insensitive to 
the confinement mechanism which is pronounced at distances 
$\sim 1$\,fm. 
Consequently, the instanton liquid model may be safely applied for 
the description of hadron properties at
the heavy quark sector too. During this applications
one can apply a systematic approach to take into account 
the nonperturbative effects in the hadron 
properties in terms of so called packing parameter of instantons
$\lambda=\rho^4/R^4$.  However,
the perturbative effects also should be carefully taken into account
during the analysis of heavy hadrons' spectra.  

\section{Heavy quark correlators with perturbative  corrections}
\label{sec:QuarkCor}

The detailed evaluation of heavy quark correlators  in the 
instanton liquid model is given in Ref.\,\cite{Musakhanov:2020hvk}. Here we quickly repeat the corresponding discussions.

As we already discussed above,  the background field due to 
instantons can be expressed in the form of simple sum 
$A(\xi)=\sum_iA_i(\xi_i)$, where  $\xi_i=(z_i,U_i,\rho_i)$ 
denotes the collective coordinates of $i^{\rm th}$ instanton. 
During calculations, one sould take into account also that the 
instanton field has a specific $A\sim 
1/g$ dependence on the strong coupling $g$. 
The normalized partition function ($Z[0]=1$) in instanton liquid model 
$Z[j]$ can be given by an approximate expression
\begin{eqnarray}
Z[j] &= \int  D\xi Da  e^{-[S_{eff}[a,A(\xi)]+(ja)]}\approx \int  D\xi e^{-\frac{1}{2}(j_\mu S_{\mu\nu}(\xi)j_\nu)},
\label{Z1}
\end{eqnarray}
which accounts the perturbative 
gluons $a_\mu$ and their corresponding sources $j_\mu$.
In obtaining the partition function in Eq.\,(\ref{Z1}), the self-interaction terms  at the order 
 of  ${\cal O}(a^3,a^4)$ are neglected and it is used the following
 definitions 
\begin{eqnarray}
 (ja)&=\int d^4x j^a_\mu(x) a^a_\mu(x), \,\,\,\cr
 (j_\mu S_{\mu\nu}(\xi)j_\nu)&=\int d^4xd^4y 
 j^a_\mu(x) S^{ab}_{\mu\nu}(x,y,\xi)j^b_\nu(y).\nonumber
\end{eqnarray}
Here $S^{ab}_{\mu\nu}(x,y,\xi)$ is a gluon propagator in the 
presence of the instanton background $A(\xi)$.
The measure of integration 
in ILM is simply given as $ D\xi=\prod_i d\xi_i=V^{-1}
\prod_i dz_{i}dU_i$ because the  instantons' sizes $\rho_i$  
due to the inter-instantons interactions are concentrated
around their average value $\bar\rho$. 
As we mentioned above,  
for the simplicity we will use $\rho_i=\bar\rho$. 

An infinitely heavy quark interacts only 
through the fourth components of 
instantons 
$A_4$ and perturbative gluon $a_4$ fields,
respectively.  Therefore, we need only $ S_{44}(\xi)$ 
components of a gluon propagator.
 Hereafter, we follow the definitions given in
 Ref.~\cite{Diakonov:1989un}, i.e. $\theta$ is inverse of
 differentiation operator $\theta^{-1}=d/dt$ and 
  $\langle t|\theta|t'\rangle
 =\theta(t-t')$ is a step-function.  For the sake of convenience,
we also use the following re-definitions of
fields $a\equiv ia_4$, $A\equiv iA_4$, 
source $j\equiv ij_4$ and 
gluon propagator $ S(\xi)\equiv S_{44}(\xi)$. 

According to these definitions and re-definitions
the heavy quark $Q$ and antiquark $\bar Q$ Lagrangians 
can be expressed as 
\begin{eqnarray}
L_Q&=Q^+(\theta^{-1}-ga-gA+...)Q,\\
L_{\bar Q}&=\bar Q^+(\theta^{-1}-g\bar a-g\bar A+...)\bar Q,
\end{eqnarray}
where the dots denote the next order in 
the inverse of heavy quark mass terms.
In terms of SU($N_c$) generators the quantities
$a$ and $\bar{a}$ are given as 
$a=a_a\lambda_a/2$ and  $\bar a_a=a_a\bar\lambda_a/2$,
where $\bar\lambda_a=-\lambda^{\rm T}_a$.\footnote{Here 
the regular superscript `T' means the operation of transposition.} 
The same rule holds for the instanton fields $A$ and $\bar A$.

During our calculations on may neglect by the virtual processes 
$Q\rightarrow QQ \bar Q$ corresponding to the heavy quark loops
which means the heavy quark determinant equals to 1. 
The functional space of heavy quarks $Q$ 
is not overlapping with the functional space of heavy 
antiquarks $\bar Q$ and, consequently,
the total functional space is a direct product of $Q$ and $\bar 
Q$ spaces.

Now the heavy quark propagator in ILM can be analyzed. 
From Eq.~(\ref{Z1}) it is seen, that the averaged heavy quark propagator ${w}$ with the account of perturbative gluon field fluctuations $a$ is given by the expression
\begin{eqnarray}
 {w}&=\int D\xi Da \exp[-S_{eff}(a,\xi)+(ja)]\left(\theta^{-1}-ga -g\sum_i A_i\right)^{-1}\cr
&=\int D\xi  \left[\int \left(\theta^{-1}-g\frac{\delta}{\delta j} -g\sum_i A_i\right)^{-1}
\exp\left\{\frac{1}{2}(jS(\xi)j)\right\}\right]_{j=0}.
\label{w1}
\end{eqnarray}
It can be easily proven that
\begin{eqnarray}
&\left[\frac{1}{ \theta^{-1}- g\frac{\delta}{\delta j} -gA(\xi)}
\exp\left(\frac{1}{2}jS(\xi)j\right)\right]_{_{j=0}}\cr
&\qquad =
\left[\exp\left(\frac{1}{2}\frac{\delta}{\delta a_a}S_{ab}(\xi)\frac{\delta}{\delta a_b}\right)
\frac{1}{ \theta^{-1}-ga -gA(\xi)}\right]_{_{a=0}}\qquad
\label{q1}
\end{eqnarray}
Furthermore, this equation can be extended to any correlator.
Consequently,  the path integral of 
heavy quark functional $F[A(\xi),a]$  in the approximations discussed above  
can be given by the following equation
\begin{eqnarray}
&\int D\xi Da \exp\left\{-S_{eff}[A(\xi),a]\right\} F[A(\xi),a]\cr
&\qquad= \int D\xi \left[\exp\left(\frac{1}{2}\frac{\delta}{\delta a_a}S_{ab}(\xi)\frac{\delta}{\delta a_b}\right) F[\xi,a] \right]_{a=0}.
\label{F}
\end{eqnarray}
Another equation similar to this equation in the absence of instanton 
background $A(\xi)=0$ and for the gluon propagator taken in 
Coulomb gauge was suggested before in Ref.~\cite{brown1979}.

As we mentioned at the end section~\ref{sec:ILM}, the systematic accounting of the nonperturbative effects in the 
ILM can be performed in terms of
the dimensionless parameter $\lambda$ by using the Pobylitsa equations~\cite{Pobylitsa:1989uq}.
The situation here is quite comfortable for the performing 
systematic analysis of instanton effects. Because $\lambda$
value is very small at the values of instanton 
parameters discussed above, 
$\lambda \sim 0.01$ (see Eq.\,(\ref{eq:SetInstanton})).

In order to take into account the perturbative OGE effects 
one should perform an expansion in terms of 
parameter $\alpha_s$. While the
behavior of $\alpha_s$  is well known at the perturbative region,
at the nonperturbative region it is not clear which value should be 
used. The pure perturbative effects at the 
leading order appear linear in $\alpha_s$.  
A systematic analysis including the both, perturbative and nonperturbative, 
effects requires a double expansion series in terms of $\alpha_s$ and $\lambda$. 
In order to perform such an analysis one may assume that 
$\alpha_s\sim \lambda^{1/2}$ which is quite reasonable according to 
the phenomenological studies.
Consequently, during the
calculations one should keep all necessary terms at the order of 
${\cal O}(\lambda)$  and
${\cal O}(\alpha_s\lambda^{1/2})$. 

\section{Gluons in  ILM}
\label{sec:ginILM}

At the approximation discussed at the end of previous section 
the gluon propagator in instanton 
medium can be represented by re-scattering series as
$$
S(\xi)=S^0+\sum_i \Delta S^i(\xi_i),\quad \Delta S^i(\xi_i)\equiv S^i(\xi_i)-S^0,
$$
where $S^0$ is free gluon propagator and  $S^i(\xi_i)$ is 
propagator of  gluon in instanton background.
The averaged value of gluon propagator $\overline{S}$ in ILM
can be found by extending the Pobylitsa's 
equation to the gluon case~\cite{Musakhanov:2017erp}
\begin{eqnarray}
\overline{S}(k)= \frac{1}{k^2+M_g^2(k)}. 
\label{glprop}
\end{eqnarray}
Consequently, the perturbative gluons are also acquire 
the momentum dependent mass and it is defined
by the following expressions
\begin{eqnarray}
M_g(k)&=M_g(0)F(k),\quad
M_g(0)=\frac{2\pi}{\rho}\left(\frac{6\lambda}{N_c^2-1}\right)
^{1/2}, 
\quad F(k)=k\rho K_1(k\rho).
\label{Mg0}
\end{eqnarray}
Here $K_1$ is a modified Bessel function of the second type.
At the typical 
values of instanton parameters $\rho=1/3\, {\rm fm},\,\,R=1\, {\rm fm}$ one can estimate 
the dynamical gluon mass at zero momentum. Its value is comes 
out $M_g(0)\simeq 358\,{\rm MeV}$ and close to the value of 
dynamical light quark mass.
One can also note, that the dynamical gluon and light quark 
masses appear 
at the order of ${\cal O}\big(\lambda^{1/2}\rho^{-1}\big)$. The  gauge invariance of the dynamical gluon mass $M_g$ was proven in Ref.~\cite{Musakhanov:2017erp}.

One may wonder that the instantons also generate
the nonperturbative gluon-gluon interactions and, in such a way,
contribute to the glueballs' properties.
The corresponding investigations in  instanton liquid model~\cite{Schafer:1994fd,Tichy:2007fk} devoted to
the $J^{PC}=0^{++}, 0^{-+},2^{++}$ glueballs, showed that the
instanton-induced forces between gluons will lead to the strong
attraction in the $0^{ ++}$ channel, to the 
strong repulsion in the 
$0^{- +}$ channel and to the absence of
short-distance effects in the $2^{ ++}$ channel. 
Consequently, applications of ILM in studies of glueballs predicted
hierarchy of the masses 
$m_{0^{++}}<m_{2^{++}}<m_{0^{-+}} $ and their 
corresponding sizes 
$r_{0^{++}}<r_{2^{++}}<r_{0^{-+}} $. These predictions were confirmed by
the lattice calculations~\cite{deForcrand:1991kc,Weingarten:1994vc,Chen:1994uw,Morningstar:1999rf,Athenodorou:2020ani,Meyer:2004jc,Meyer:2004gx}.
At typical values of ILM 
parameters $\rho=1/3$~fm and  $R=1$~fm 
there were found~\cite{Schafer:1994fd}, that the mass of $0^{++}$ 
glueball $m_{0^{++}}=1.4\pm 0.2$~GeV and  its size  
$r_{0^{++}}\approx 0.2$~fm in a nice correspondence with
the lattice calculations~\cite{deForcrand:1991kc,Weingarten:1994vc,Chen:1994uw}.
Further studies of the $0^{++}$ glueball in ILM~\cite{Tichy:2007fk} gave 
$m_{0^{++}}=1.29\, -\, 1.42$~GeV, which was also 
in a good agreement with the lattice 
results~\cite{Meyer:2004jc,Meyer:2004gx}.
 
Main conclusion of the works we discussed above
was that the origin of $0^{++}$  glueball is mostly provided by the short-sized 
nonperturbative fluctuations (instantons), rather than the confining forces. In a quick summary, one may conclude that 
ILM provides the consistent framework for describing the
gluon and  the lowest state glueball's properties.

\section{Heavy quark propagator in ILM}
\label{sec:HQP}

Hereafter, we concentrate on the propereties of heavy quarks
and the heavy quark systems in the framework of instanton 
liquid model. Let us first discuss a single heavy quark properties in 
 ILM by estimating the corresponding effects from  
perturbative and nonperturbative regions as it was done in Ref.\,\cite{Musakhanov:2020hvk}.

An averaged infinitely heavy quark $Q$ propagator in ILM 
according to Eqs.(\ref{w1})-(\ref{q1}) is given as
\begin{eqnarray}
\label{wQ}
 {w}=\!\left.
\!\int\!\! D\xi \exp\left[\frac{1}{2}\!
\left(\frac{\delta}{\delta a}S(\xi)\frac{\delta}{\delta a}\right)\!\right] \!
\frac{1}{\theta^{-1}-ga -gA(\xi)}  \right|_{a=0}\!\!\!
\end{eqnarray}
where we have used the definition
\begin{equation}
\left(\frac{\delta}{\delta a}S(\xi)\frac{\delta}{\delta a}\right)=
\int d^4yd^4z
\frac{\delta}{\delta a_a(y)}S_{ab}(\xi,y,z)\frac{\delta}{\delta a_b(z)}.
\end{equation}
The details of systematic analysis of the heavy quark propagator is discussed in Appendix of Ref.\,\cite{Musakhanov:2020hvk}.  
From there one can see that in the instanton liquid model the heavy quark propagator with perturbative corrections 
can be written as
\begin{equation}
 {w}=\int D\xi\left[\theta^{-1} -\sum_i \left( g A_i(\xi_i) -g^2\left( \Delta S^i(\xi_i)\theta\right)\right)\right]^{-1}.
\label{w11}
\end{equation}
Here the last term in the denominator means the heavy quark mass operator of the order ${\cal O}(\alpha_s\lambda^{1/2})$.
Heavy quark propagator Eq.~(\ref{w11}) and its $g\rightarrow 0$ limit expression have the similar structures according to their dependencies on the instanton collective coordinates.
One can now extend Pobylitsa equation in Ref.~\cite{Diakonov:1989un} and the corresponding extension in 
the approximation ${\cal O}(\lambda,\alpha_s\lambda^{1/2})$ has form
\begin{eqnarray}
 {w}^{-1}&=&
\theta^{-1}-\sum_i \int d\xi_i 
\theta^{-1}\left(\frac{1}{ \theta^{-1} -g A_i(\xi_i)}-\theta\right)\theta^{-1}-g^2 \left((\bar S-S^0) \theta\right).
\label{wQ-1}
\end{eqnarray}
In the last term the averaged 
gluon propagator $\bar S$ is given by Eq.~(\ref{glprop}).
In such a way, the second term 
in the right side of Eq.~(\ref{w-1}) leads to the ILM heavy 
quark mass shift $\Delta M_Q^{\rm dir}$ with the corresponding 
order  ${\cal O}(\lambda)$ while the third one is ILM modified  
perturbative gluon contribution to the heavy quark mass 
$\Delta M_Q^{\rm pert}$ with order of  ${\cal O}(\alpha_s\lambda^{1/2})$, respectively. 
We note that the direct mass contribution to the quark mass in 
instanton background $\Delta M_Q^{\rm pert}$ was calculated
first in Ref.~\cite{Diakonov:1989un}.
At the typical values of parameters $N_c=3$, $\alpha_s= 0.3,\,
\rho=1/3\, {\rm fm}$, $R=1\, {\rm fm}$ one can estimate
 $$
\Delta M_Q^{\rm pert}\le 
\frac{2}{N_c}\,\alpha_s M_g(0)
\sim  \Delta M_Q^{\rm dir}\simeq 70\, {\rm MeV}.
$$
 This estimation is in accordance with the above made assumptions 
 ${\cal O}(\alpha_s\lambda^{1/2})\sim {\cal O}(\lambda)$ and
shows that the instanton-perturbative gluon interaction 
changes the perturbative gluon corrections.

\section{Phenomenological potential models}
\label{sec:PotMod}

Further, we concentrate on the properties of quarkonium 
$Q\bar Q$ 
(a colorless system consisting a heavy quark $Q$ and another 
heavy antiquark $\bar Q$) and will discuss the contributions 
from the nonperturbative dynamics in describing their 
properties. As an example, we 
analyze the nonperturbative effects in charmonium spectrum.
The non-relativistic quantum-mechanical potential 
approaches can be readily applied for describing the charmonum
spectrum~\cite{Brambilla:2010cs,Eichten:1974af, Eichten:1978tg, Eichten:2007qx,Voloshin:2007dx}. 

In a standard approach there are
basically two main contributions to the heavy-quark potential.
It is so-called
Cornell potential~\cite{Eichten:1974af}, which has a nature of 
Coulomb-like attractive part at short 
distances and linear confining part at long distances.
The form of potential is given as
\begin{equation}
V_{\rm Cornell}(r)=\frac{\kappa}{r}+\sigma r\,,
\label{cornell}
\end{equation}
where the Coulomb coupling $\kappa<0$ and 
the string constant $\sigma>0$.
The Coulomb-like potential originates from
one-gluon exchange (OGE) between a heavy quark $Q$ and a heavy
anti-quark $\bar{Q}$~\cite{Susskind:1976pi, Appelquist:1977tw, 
  Appelquist:1977es, Fischler:1977yf}.
This potential can be calculated based on 
perturbative quantum chromodynamics (pQCD)
 and at the leading order on the strong 
coupling constant $\alpha_s=g^2/(4\pi)$, one reproduces 
the constant $\kappa =-(4/3)\alpha_s$ in 
the first term in Eq.\,(\ref{cornell}).
Note, that the static Coulmob-like potential was scrutinized already 
to higher-order corrections from pQCD~\cite{Peter:1996ig, 
  Peter:1997me, Schroder:1998vy, Smirnov:2009fh,
  Anzai:2009tm}. By nature of pQCD, the Coulomb-like interactions 
  are supposed to govern the short-range physics of charmonia. 
  
At large distances the strength of the Coulomb-like interaction 
decreases but the presence of the quark-confining potential will 
increase the strength of total interaction.
In such a way quarks inside of a charmonium is 
confined~\cite{Wilson:1974sk}. The heavy-quark potential for the 
quark
confinement can be obtained at least phenomenologically from the
Wilson loop, which rises linearly at large
distances~\cite{Eichten:1974af, Eichten:1978tg}.   
Actually, there are also different type of potentials are in use.
For example, the harmonic oscillator type $\sim r^2$ or the 
logarithmic $\sim \ln (r)$ dependencies at the confining region.
From other side, the lattice QCD calculations showed the linear $\sim r$ dependence of the full potential
at large distances~(see, e.g. Ref.\,\cite{Bali:2000gf})
supporting in such a way the Cornell-type form of
potentials. Actually, all these potential models 
with the different confining forms
reasonable well match 
with the data. The reason behind is that they do not much affect at 
short distances where we have the sensitive probes
to the form of confining potential.  
This kind of common 
behavior of different confining potentials at the short 
distances is also partial reason for considering 
the Coulomb-coupling  $\kappa$ as a pure 
phenomenological parameter.
 
One can further try to develop the potential 
approach and improve the description
of data by taking into account the 
relativistic and perturbative corrections on the strong 
coupling constant $\alpha_s$ of QCD~\cite{Brambilla:2009bi,Mateu:2018zym}.  From 
other side,  one may also expect 
that the non-perturbative effects on the heavy hadron properties 
in the instanton vacuum could be substantial.
In the following we will consider such effects in the 
properties of charmonium. 
We mainly concentrate on the non-perturbative 
effects, but simultaneously consider
the perturbative gluon contributions too. 

\section{Instanton contributions to the heavy quark potential}
\label{sec:IconPot}

The detailed calculation of $Q\bar Q$ correlator in the 
instanton vacuum and obtaining the corresponding 
interaction potential based on Wilson-loop formalism 
 is discussed in Ref.\,\cite{Musakhanov:2020hvk}.
Here we present the final form of instanton contributions 
and discuss the corresponding effects.

\subsection{Direct instanton induced singlet potential in ILM}
\label{subsec:dir}

The direct instanton induced 
potential $V_{\rm dir}(r)$ can be evaluated 
by repeating the calculations presented in Ref.~\cite{Diakonov:1989un}.
It has the following final form
\begin{equation}
V_{\rm dir}(r)=\frac{4\pi\lambda}{N_c\rho}\,
{\cal I}_{\rm dir}\left(\frac{r}{\rho}\right),
\label{VIdir}
\end{equation}
where ${\cal I}_{\rm dir}(x)$ - dimensionless integral expressed as
\begin{eqnarray}
{\cal I}_{\rm dir}(x)&=&\int_0^\infty y^2 dy \int_{-1}^1 dt\bigg[1-\cos\left(\frac{\pi y}{\sqrt{y^2+1}}\right)
\cos\left(\pi\sqrt{\frac{y^2+x^2+2xyt}{y^2+x^2+2xyt+1}}\right)\cr
&-&\frac{y+xt}{\sqrt{y^2+x^2+2xyt}}\sin\left(\frac{\pi y}{\sqrt{y^2+1}}\right)\sin\left(\pi\sqrt{\frac{y^2+x^2+2xyt}{y^2+x^2+2xyt+1}}\right)\bigg].
\label{Idirnumer}
\end{eqnarray}
At the small distances ($x\ll 1$), ${\cal I}_{\rm dir}(x)$ can be evaluated
analytically and one has the potential
\begin{eqnarray}
V_{\rm dir}(r)&\simeq &\frac{4\pi\lambda}{N_c\rho}\,\left\{\frac{\pi^2}{3}\left[\frac{\pi}{16}-J_1(2\pi)\right]\frac{r^2}{\rho^2}-\pi\left[\frac{\pi^2(438+7\pi^2)}{30720}+\frac{J_2(2\pi)}{80}\right]\frac{r^4}{\rho^4}
\right\}\!,\qquad
\end{eqnarray}  
in terms of the Bessel functions $J_{n}$. At the large values of the 
$Q\bar{Q}$ inter-distance ($r\gg \rho$),
the potential has the form 
\begin{equation}
V_{\rm dir}(r)\simeq 2\Delta M_Q^{\rm dir}-
\frac{2\pi^3\lambda}{N_c r}.
\label{VNPlargex}
\end{equation}
One can see that the direct instanton potential mainly
contributes at perturbative region as an overall shift 
in the spectrum of quarkonium states.

\subsection{Perturbative one-gluon exchange singlet potential in ILM}
\label{subsec:1gluon}

\begin{figure}[h]
\begin{center}
\includegraphics[width=7cm,angle=0]{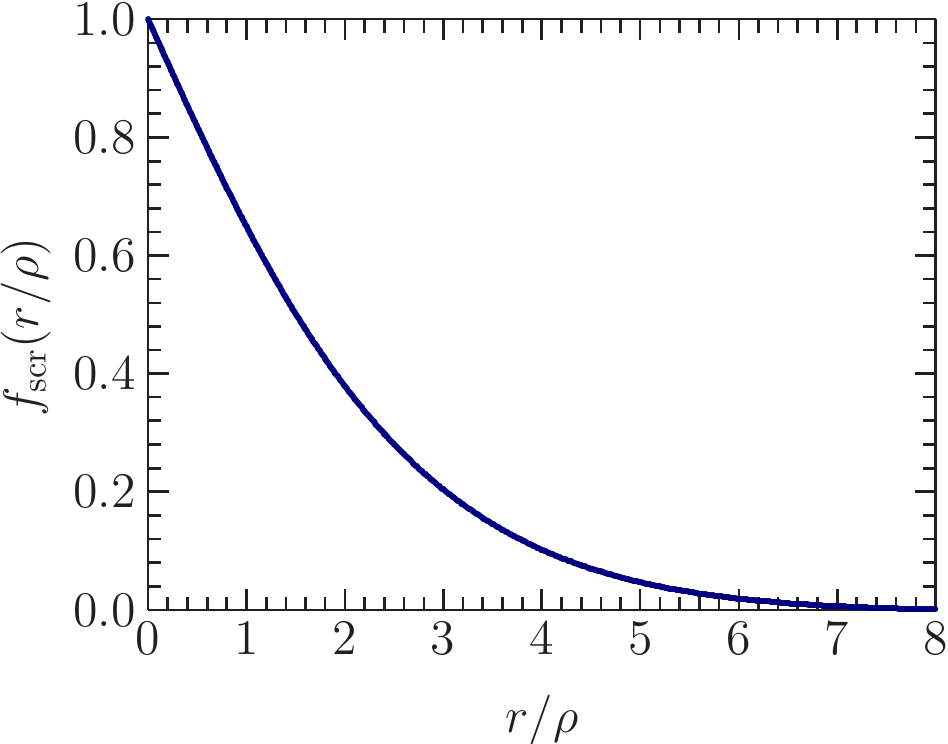}
\end{center}
\caption{Dependence of screening function 
$f_{\rm scr}$ on the dimensionless variable 
$r/\rho$ at ILM parameters
$ R=1~{\rm fm}$ and $\rho=1/3~{\rm fm}$.}
\label{fscreen}
\end{figure}

Calculation of the perturbative one-gluon-exchange potential
 in the presence of instanton background gives the following 
final form\,\cite{Musakhanov:2020hvk}
\begin{equation}
V_{\rm pert}(r)=-\frac{4}{3}\,
g^2\int\frac{d^3q}{(2\pi)^3}\frac{e^{i\vec q\cdot\vec r}}{q^2+M_g^2(q)},
\label{1gl}
\end{equation}
where $M_g(q)$ is given by Eq.~(\ref{Mg0}).
One can see that, the dynamical mass generation of gluons
plays the screening effect at perturbative region.
To see this effect explicitly we rewrite the potential\,(\ref{1gl})
after the integration over angular variables
\begin{eqnarray}
\label{Vpertup}
V_{\rm pert}(r)&=-\frac{4\alpha_s}{3r}f_{\rm scr}\left(\frac{r}{\rho}\right),\\
f_{\rm scr}(x)&=1-\frac{2x}{\pi}
\int_0^\infty 
{\rm d}y~j_0(xy)
\frac{3\pi^2
\lambda{ K}_1^2(y)}{1+3\pi^2
\lambda{ K}_1^2(y)},
\label{Vpert}
\end{eqnarray}
where $f_{\rm scr}(x)$ plays the role of screening function.
It is presented in Fig.~\ref{fscreen}.
Naturally, in the absence of instantons ($\lambda=0$) 
one restores the standard perturbative OGE potential. 
At small values of $r\rightarrow 0$ the screened OGE potential 
can be approximated by a Yukawa-type potential
\begin{eqnarray}
V_{\rm pert}(r)&\simeq  -\frac{4\alpha_s}{3r}\exp(-M_Y r)\,,\quad
M_Y =\frac{2}{\pi\rho}\,\int_0^\infty dy\,\frac{3\pi^2
\lambda{ K}_1^2(y)}{1+3\pi^2
\lambda{ K}_1^2(y)}\,,
\label{VQQpersmallx}
\end{eqnarray}
where $M_Y< M_g(0)\simeq 358\,{\rm MeV}$
at the given values of $\rho=1/3$~fm, $R=1$~fm. 

At large distances $r>\rho$ 
the potential $V_{\rm pert}(r)$  is not long-ranged anymore 
and quickly goes to zero. In such a way, at large distances 
the instanton medium produces the screening effect 
in the one gluon exchange perturbative potential.

\section{Order of instanton effects}
\label{sec:InsEffects}

For the quick estimation of the order of instanton effects, 
one may ignore the spin splitting 
effects in the charmonium spectra and concentrate only 
on some low lying S-wave states. 
It is known that, although the direct instanton potential 
shows the linear behavior at small and intermediate distances
it is flattened at perturbative region approaching in such a way
the constant value (see Eq.\,(\ref{VNPlargex})).
Therefore, the instantons cannot provide confinement
and the confining potential should be added into the model
in a phenomenological way.

Further, one defines the full
central $Q\bar{Q}$ potential which includes all possible instanton effects in the following form\,\cite{Musakhanov:2020hvk} 
\begin{equation}
V(r)=\sigma r + V_{\rm pert}(r)+V_{\rm dir}(r),
\end{equation}
which supplies the confinement phenomenon at large distances.
This potential leads to the standard Cornell's potential
Eq.\,(\ref{cornell}) in the absence of instanton effects.
The order of the instanton effects may be estimated 
by applying a time independent perturbation 
approach and comparing the perturbative calculations 
with the full variational calculations. For that purpose, 
one can divide the Hamiltonian into two parts 
\begin{equation}
H=H_0+\tilde H.
\label{FullH}
\end{equation} 
Here $H_0$ is Hamiltonian of the Cornell's model and 
$\tilde H$ is the perturbative part of Hamiltonian due 
to instanton contributions. The corresponding 
parts of the full Hamiltonian is defined as
\begin{eqnarray}
H_0&=-\frac{1}{m_Q}\vec\nabla^2+V_{\rm Cornell},\cr
\tilde H&= V-V_{\rm Cornell} \equiv V_{\rm dir}+V_{\rm scr},\cr
 V_{\rm scr}&=-\frac{4\alpha_s}{3}
(f_{\rm scr}-1).
\end{eqnarray}
The details of the full variational calculations can 
be found in Ref.\,\cite{Yakhshiev:2018juj}.

The results of  calculations are presented in 
Table~\ref{Table2}.  As an example 
of the Cornell's model parameters, it is chosen an approximated  
parameter set MWOI presented in Table\,I of
Ref.\,\cite{Yakhshiev:2018juj}. 
\begin{table}[h]
\caption{The results of full variational calculations.
 It is considered 
only some of the  S-wave states corresponding to 
the charmonium states (Spin dependent parts of 
interactions are not included).  The chosen as
as $m_Q=1275$\,MeV, $\sigma=0.17$\,GeV$^2$, $\alpha
=0.2$. The ILM parameter are given in Eq.\,(\ref{eq:SetInstanton})}.
\footnotesize
\begin{center}
\begin{tabular}{@{}ccccc}
\hline 
$n$& $V_{\rm Cornell}$ & 
$V_{\rm Cornell}+V_{\rm scr}$& $V_{\rm Cornell}+V_{\rm dir}$& $V$\\
\hline
 1& 3069 &3129 &3111 &3172\\
 2& 3611 &3664 &3682 &3736\\
  3 & 4035 &4079 &4119 &4163\\
  4 & 4405 &4443&4496&4534\\
   \hline
\end{tabular}
\end{center}
\label{Table2}
\end{table}
For comparison, in Table\,\ref{Table2}
we present the results for 
Cornell's potential,``Cornell + instanton'' potentials which have 
the nature of instanton contributions from the different regions
and also for the full potential which takes into account all 
possible instanton effects from the different regions. 
The both potentials $V_{\rm scr}$ and 
$V_{\rm dir}$ are positively defined and, therefore, 
give the positive contributions to  the whole spectrum. 
This is seen from the corresponding results 
in Table\,\ref{Table2}.

The instanton contributions are not big but 
they are not negligible too.   In order to 
understand this situation better one can calculate the 
first order perturbative corrections to the Cornell's model 
results considering the instanton effects as the small 
perturbations. 

The comparisons of the corresponding 
perturbative and fully variational 
calculations are shown in Table~\ref{Table3}.
\begin{table}[h]
\caption{The perturbative vs full variational calculations.
The first column is radial exitations, 2-4 columns are 
the first order perturbative corrections, 5-7 columns are 
the corresponding differences of variational calculations with and without instanton 
generated potentials,
respectively (see explanations in the text).
The parameters and other definitions are same 
as in the Table\,\ref{Table2}.}
\footnotesize
\begin{center}
\begin{tabular}{@{}cccc|ccc}
\hline 
&\multicolumn{3}{c|}{First order perturbative corrections}&
\multicolumn{3}{c}{The corresponding variational calculations}\\
$n$ & $V_{\rm scr}$&  $V_{\rm dir}$& $V_{\rm scr}+V_{\rm dir}$ &``$V_{\rm scr}$''&  ``$V_{\rm dir}$''& ``$V_{\rm scr}+V_{\rm dir}$''\\
\hline
1&60.124&44.305&104.430&60.119&42.439&102.611\\
2&52.826&72.224&125.050&52.707&71.438&124.651\\
3&43.864&84.342&128.206&43.743&83.873&127.954\\
4&38.247&91.518&129.765&38.172&91.193&129.561\\
   \hline
\end{tabular}
\end{center}
\label{Table3}
\end{table}
In the left-half of the table, it is presented 
the first order perturbative corrections due to instantons
(see $\tilde H$ in Eq.\,(\ref{FullH}))
calculated on a basis of Cornell's
model wave functions corresponding to the Hamiltonian $H_0$.  
On the right-half of the table, it is presented
the corresponding differences of variational calculations with and without instanton 
generated potentials. For example, 
``$V_{\rm scr}$" means the difference between the results 
of the potential models, ``$V_{\rm Cornell}+V_{\rm scr}$'' 
and ``$V_{\rm Cornell}$", obtained by means the variational 
calculations (The corresponding results  are 
presented in Table\,\ref{Table2}.). 
It should be compared with the 
first order perturbative corrections corresponding to the 
perturbation potential $V_{\rm scr}$. 
One can see, that the instanton effects can be considered as 
the first order 
perturbative corrections to the spectrum.

When the value of $\alpha_s$ is changed, the general 
picture will not change if one concentrates to the order of 
instanton contributions, i.e. they still remain as the 
first order perturbative corrections. 
The relative sizes of all possible instanton effects 
in comparison with the results corresponding to the 
Cornell's model results found to
be few percents depending on the parameters of 
instanton liquid model and the excitation state. 

\section{Heavy and light quarks in the instanton vacuum}
\label{sec:hlqinILM}

Now let us discuss the systems containing heavy and light 
quarks. While the instantons govern the light quark physics
completely, in the heavy quarks sector they may only 
affect the heavy quark  mass and heavy quark-quark 
interactions\,\cite{Diakonov:1989un,Turimov:2016adx,Yakhshiev:2018juj,Musakhanov:2020hvk,Chernyshev:1995gj}
as we discussed above. 
In the heavy-light system, the instantons generate heavy-light quark interaction terms 
which are responsible for the corresponding 
chiral symmetry breaking effects\,\cite{Musakhanov:2014fya}. 

\subsection{Light quarks in ILM}

As a starting point, one can represent the light quark determinant 
${\rm Det}$ as a product of
the low and high frequency parts ${\rm Det}={\rm Det}_{{\rm high}}\cdot{\rm Det}_{{\rm low}}$. Here  ${\rm Det}_{{\rm high}}$ gets the contribution from the fermion modes with Dirac eigenvalues at 
the interval from arbitrary 
$M_{1}$ to the Pauli--Villars mass $M$  and 
${\rm Det}_{{\rm low}}$
 accounts the eigenvalues less than $M_{1}$. In general, 
 the product of these
determinants is independent on the scale of $M_{1}$. However,
one can calculate both of them only approximately. 
Calculations show, that there is a week 
$M_{1}$ dependence of ${\rm Det}$ as the
product in the wide range of $M_{1}$. This serves as
a check of the approximations used in Ref.~\cite{Diakonov:1995qy}. 
The high-frequency part ${\rm Det}_{{\rm high}}$ can be written as a
product of the determinants in the field of individual instantons.
The low-frequency part ${\rm Det}_{{\rm low}}$ is influenced by 
the whole ensemble of instantons and approximately would be 
that which accounts only the zero modes~\cite{Diakonov:1995qy}.

\begin{figure}[!ht]\center
\tikz{\begin{feynman}
\vertex (i1){};
\vertex(q1) [below=1cm of i1]{$q_1$};
\vertex(m1) [below=0.5cm of q1];
\vertex(n1) [below=0.5cm of m1];
\vertex(o1) [below=0.5cm of n1]{$q_{N_f}$};
\vertex(c)[blob] [right=2.2cm of q1]{$\rm I$};
\vertex(q2) [right=2.2cm of c]{$q_1$};
\vertex(n2) [above=0.5cm of q2];
\vertex(m2) [above=0.5cm of n2];
\vertex(o2) [above=0.5cm of m2]{$q_{N_f}$};
\diagram*{
(q1) --[fermion] (c) --[fermion] (q2),
(o1) --[fermion] (c) --[fermion](o2);
};\end{feynman}}
\caption{Instanton generated $N_f$ light $q$ quarks interaction.}
		\label{figq}
	\end{figure}
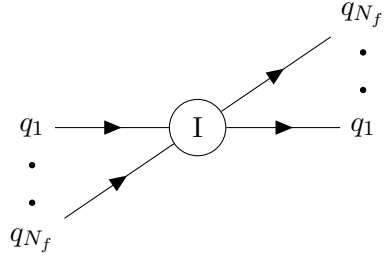

Again, the instanton background field is assumed to be as the
superposition of $N_{+}$ instantons and $N_{-}$ antiinstantons 
(see reviews~\cite{Diakonov:2002fq,Schafer:1996wv}).
By summing the light quarks-instantons re-scattering series which
leads to the total light quark propagator $S$ and making further few
steps, one can can get the fermionized representation of  the
low-frequency light quark determinant in the presence of the quark sources. 
It has the form~\cite{Goeke:2007bj,Goeke:2007nc,Goeke:2010hm,Musakhanov:1996qf,Salvo:1997nf,Musakhanov:1998wp,
Musakhanov:2002vu,Musakhanov:2002xa,Kim:2004hd,Kim:2005jc} 
\begin{eqnarray}
\label{part-func}
&{\rm Det}_{\rm low}&\exp(-\eta^{+}S\eta)=\int\prod_{f}D\psi_{f}D\psi_{f}^{\dagger}\cr
&&\times\exp\int\sum_{f}\left(\psi_{f}^{\dagger}(\hat{p}\,+\, im_{f})\psi_{f}+\psi_{f}^{\dagger}\eta_{f}+\eta_{f}^{+}\psi_{f}\right)
\prod_{f} \prod_{\pm}^{N_{\pm}}V_{\pm,f}[\psi^{\dagger},\psi],
\end{eqnarray}
 where $\psi$ is quark field, $\eta$ is the corresponding 
 quark source, $S=(\hat p-\hat A+im)^{-1}$ is quark propagator in the instanton medium and
 \begin{eqnarray}
V_{\pm,f}[\psi^{\dagger},\psi] = i\int
   d^{4}x\left(\psi_{f}^{\dagger}(x)\,
   \hat{p}\Phi_{\pm,0}(x;\xi_{\pm})\right)\int
   d^{4}y\left(\Phi_{\pm,0}^{\dagger}(y;\xi_{\pm})(\hat{p}\,\psi_{f}(y)\right), 
\label{V}
\end{eqnarray}
is instanton generated $N_f$ light quarks interaction represented in Fig.~\ref{figq}.

The averaging over the collective coordinates $\xi_{i,\pm}$ of
${\rm Det}_{\rm low}\exp(-\eta^{+}S\eta)$  is a rather simple procedure, 
since the low density of the instanton medium
($\lambda\sim0.01$) allows to average
over positions and orientations of the individual instantons 
independently from each other. This process
leads to the light quark partition function $Z[\eta,\eta^+]$. 
At the single flavor sector $N_f=1$ and for the equal number 
of instantons and anti-instantons $N_\pm=N/2$, using Eq.\,(\ref{part-func})  one can 
get the exact form of partition function  
\begin{eqnarray}
Z[\eta,\eta^+]&=&\exp\{-\eta^+\left[\hat p \,+\, i(m+M(p))\right]^{-1}\eta\}\cr
&&\times\exp\left\{\Tr\ln\left[\hat p \,+\, i(m+M(p))\right]+N\ln\frac{N/2}{\lambda}-N\right\},
\label{Z}
\\
&&N=\Tr\frac{iM(p)}{\hat p \,+\, i(m+M(p))},\qquad M(p)=\frac{\lambda}{N_c}(2\pi\rho F(p))^2.
\label{M}
\end{eqnarray}
Here $\Tr(...)=\tr_{c,f,D}\int d^4x\langle x|(...)|x\rangle$ is a 
whole functional trace, 
$\tr_{c,f,D}$ is the trace over color, flavor and Dirac indexes and the form-factor has form
\begin{equation}
F(p) = 2z \left(I_0 (z) K_1 (z)- I_1 (z) K_0 (z)
-\frac{1}{z} I_1 (z) K_1 (z)\right).
\end{equation}
Here $I_0$, $I_1$ and $K_0$, $K_1$ are the modified Bessel
functions of the first and second kind, respectively,
$z=p\rho/2$, the form-factor  $F(p)$ is
obtained by Fourier-transform of the
zero-mode and the dynamical quark mass $M(p)$
is defined in Eq.\,(\ref{M}). One can see, that due 
to the instanton induced interactions the quark mass becomes 
momentum dependent and, in such a way, generates
 the constituent quark mass.

 At $N_f >1$, the saddle-point approximation (without meson 
 loop contributions)  gives the generating functional
  $Z[\eta_f,\eta_f^+]$ which 
 has a similar form with that which is expressed one in Eq.\,(\ref{Z}).
 
 \subsection{Heavy quark-light quarks interactions in ILM}
 
\begin{figure}[!ht]\center
\tikz{\begin{feynman}
\vertex (i1){$Q$};
\vertex(q1) [below=1cm of i1]{$q_1$};
\vertex(m1) [below=0.5cm of q1];
\vertex(n1) [below=0.5cm of m1];
\vertex(o1) [below=0.5cm of n1]{$q_{N_f}$};
\vertex(c)[blob] [right=2.2cm of q1]{$\rm I$};
\vertex(q2) [right=2.2cm of c]{$q_1$};
\vertex(i2) [below=1cm of q2]{$Q$};
\vertex(n2) [above=0.5cm of q2];
\vertex(m2) [above=0.5cm of n2];
\vertex(o2) [above=0.5cm of m2]{$q_{N_f}$};
\diagram*{
(i1) --[fermion,line width=1.5pt] (c) --[fermion,line width=1.5pt] (i2),
(q1) --[fermion] (c) --[fermion] (q2),
(o1) --[fermion] (c) --[fermion](o2);
};\end{feynman}}
\caption{Instanton generated heavy $Q$ and $N_f$ light $q$ quarks interaction.}
		\label{figQq}
	\end{figure}
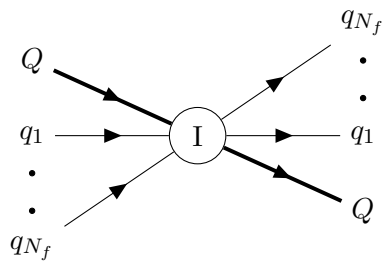

\begin{figure}[!ht]\center
\tikz{\begin{feynman}
\vertex (i1){$Q$};
\vertex(i2)[blob] [right=2cm of i1]{$\rm I$};
\vertex(i3) [right=2cm of i2]{$Q$};
\vertex(o1) [above=1cm of i3]{$\pi$};
\vertex(o2) [below=1cm of i3]{$\pi$};
\diagram*{
(i1) --[fermion,line width=1.5pt] (i2) --[fermion,line width=1.5pt] (i3),
(i2) --[scalar] (o1),
(i2) --[scalar] (o2);
};\end{feynman}}
\caption{Instanton generated emission of pions by heavy $Q$ quark.}
		\label{figQpi}
	\end{figure}
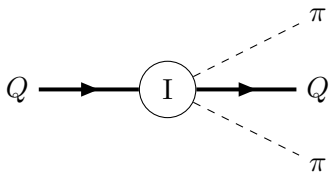

Let us first consider the simplest heavy quark correlator, i.e. the heavy quark propagator in ILM in the presence of light quarks~\cite{Musakhanov2018}. 
 We will
extend the equation for the heavy quark propagator $w$ in the instanton
media previously derived in
Refs.~\cite{Diakonov:1989un,Pobylitsa:1989uq}, by accounting in the measure of light quarks' determinant  
 as: 
\begin{eqnarray}
\label{w}
w=\frac{1}{Z}&\!\int \prod_{f}D\psi_{f}D\psi_{f}^{\dagger}\exp\int\sum_{f}\left(\psi_{f}^{\dagger}(\hat{p}\,+\, im_{f})\psi_{f}\right)
\cr
&\times \prod_{\pm}^{N_{\pm}}\langle{\prod_{f}V_{\pm,f}[\psi^{\dagger},\psi]}\rangle w[\psi,\psi^\dagger],\\
&
\langle{\prod_{f}V_{\pm,f}[\psi^{\dagger},\psi]}
\rangle\,\equiv\,\int d\xi_\pm\prod_{f}V_{\pm,f}[\psi^{\dagger} ,\psi ],
\\
w[\psi,\psi^\dagger]&
=\left\{\prod_{\pm}^{N_{\pm}}\langle\prod_{f}V_{\pm,f}[\psi^{\dagger},\psi]\rangle
\right\}^{-1}\!\!\!
\int\prod_{\pm}^{N_{\pm}} d\xi_\pm
\left\{\prod_{\pm}^{N_{\pm}}V_{\pm,f}[\psi^{\dagger} ,\psi ]\right\}\frac{1}{\theta^{-1}-\sum_i A_i}.
\end{eqnarray}
The measure of the integration over  $\xi_\pm$ in the Eq.~(\ref{w}) with and without light quark factor 
$\prod_{f}V_{\pm,f}[\psi^{\dagger},\psi]$ has the same structure as a product of independent integrations over the instanton collective coordinates  $\xi_\pm$. Then, we may extend the derivation of the Pobylitca equations~\cite{Diakonov:1989un,Pobylitsa:1989uq} 
 and solve them at the instantons low density approximation as
\begin{eqnarray}
w^{-1}[\psi,\psi^\dagger]=\theta^{-1} - \frac{N}{2}\sum_\pm \frac{1}{<{\prod_{f}V_{\pm,f}[\psi^{\dagger},\psi]}>}\Delta_{H,\pm}[\psi^{\dagger},\psi ] + O(N^2/V^2),
\label{w-1}
\end{eqnarray}
where, defining the heavy quark propagator in the
single (anti)instanton field as $w_\pm=(\theta^{-1}-A_\pm)^{-1}$, we have
\begin{eqnarray}
\Delta_{H,\pm}[\psi^{\dagger},\psi ]=\int d\zeta_\pm\prod_{f}V_{\pm,f}[\psi^{\dagger},\psi]\theta^{-1}(w_\pm-\theta)\theta^{-1}.
\label{Delta}
\end{eqnarray}
The last expression represents the interactions of heavy $Q$ and $N_f$ light quarks $q$ generated by instantons as (see Fig.~\ref{figQq})
\begin{eqnarray}
S_{Qq}= - \lambda\sum_\pm\int Q^\dagger\Delta_{H,\pm}[\psi^{\dagger},\psi ]Q .
\end{eqnarray}
It is obvious that the light quarks are emitted in colorless  
or in gluon-like colorful states. Let us consider $N_f=2$ colorless 
light quarks states. These states are represented by mesons and we 
consider the lightest one -- pions, as shown 
in Fig.\,\ref{figQpi}. 
Technically, it can be done by bosonization method and  the corresponding amplitude of the emission of pions 
will have form $Q\rightarrow Q \pi \pi$~\cite{Musakhanov2018}  
\begin{eqnarray} 
\label{AQpi}
A_{Q\pi}&=&F^2_{\pi Q}\int d^4x\,
\tr_f\partial_\mu U(x) \partial_\mu U^+(x) \frac{\exp(-ipx)}{(2\pi)^4} \frac{1}{N_c}\tr_c T_\pm(\vec p,0,0),
\\
&&
\frac{1}{N_c}\tr_c T_\pm(\vec p,0,0)=\frac{2}{N_c}J_0(|\vec p\,|\rho)=-\Delta M R^4\frac{J_0(|\vec p\,|\rho)}{J_0(0)}.
\label{AQpions}
\end{eqnarray}
Here 
\begin{eqnarray} 
F^2_{\pi Q} =2N_c\int\frac{d^4p}{(2\pi)^4}\frac{M^2(p)p^2}{(p^2+M^2(p))^3 }
\label{FpiQ}
\end{eqnarray}
and $M(p)$ is the dynamical light quark mass. At the values of ILM 
parameters $\rho=1/3\,{\rm fm},\,\,\,R=1\,{\rm fm}$ 
one can get $F_{\pi Q}=0.6 F_{\pi}$. Here  the pion decay constant 
value $F_\pi=72\,{\rm MeV}$ is obtained from the expression
\begin{eqnarray}
F^2_{\pi } =4N_c\int\frac{d^4p}{(2\pi)^4}\frac{M^2(p)}{(p^2+M^2(p))^2 }.
\end{eqnarray}
The integration over $x$ in Eq.~(\ref{AQpi}) gives the energy-momentum conservation in the form of $\delta$-function.

\subsection{Heavy quarkonium and light quarks interactions in ILM}
\label{QQq}

Now let us consider ILM $Q\bar Q$ correlator with the account of light quarks, defined as
\begin{eqnarray}
\langle 0|T&&(Q^\dagger(t_2',\vec x_2)\bar Q^\dagger(t_1',\vec x_1))\bar Q(t_2,\vec x_2)Q(t_1,\vec x_1)|0\rangle
=\frac{1}{Z}\int D\psi_f D\psi_f^{\dagger}\cr
&&\times
\left\{\prod_{\pm}^{N_{\pm}}\langle  V_{\pm}[\psi^{\dagger} ,\psi ]\rangle\right\}
\exp\left(\int\sum_f\psi_f^{\dagger}(\hat p+im_f )\psi_f\right) \cr
&&\times\langle t_1',\vec x_1; t_2',\vec x_2|W[\psi,\psi^\dagger]|t_1,\vec x_1; t_2,\vec x_2\rangle .
\label{QQcorrelator}
\end{eqnarray} 
Here
\begin{eqnarray}
\langle t_1',\vec x_1;&& t_2',\vec x_2|W[\psi,\psi^\dagger]|t_1,\vec x_1; t_2,\vec x_2\rangle
=\left\{\prod_{\pm}^{N_{\pm}}\langle  V_{\pm}[\psi^{\dagger} ,\psi ]\rangle\right\}^{-1}\cr
&&\times \int D\zeta
\left\{\prod_{\pm}^{N_{\pm}}\prod_f V_{\pm,f}[\psi_f^{\dagger} ,\psi_f ]\right\}
\langle t_1',\vec x_1|\left(\theta^{-1}-\sum_i A^{(1)}_i\right)^{-1}|t_1,\vec x_1\rangle\cr
&&\times
\langle t_2',\vec x_2|\left(\theta^{-1}-\sum_i\bar A^{(2)}_i\right)^{-1}|t_2,\vec x_2\rangle\cr
&& 
=\left\{\prod_{\pm}^{N_{\pm}}\langle  V_{\pm}[\psi^{\dagger} ,\psi ]\rangle\right\}^{-1}\int D\zeta
\left\{\prod_{\pm}^{N_{\pm}}\prod_f V_{\pm,f}[\psi_f^{\dagger} ,\psi_f ]\right\}\cr
&&\times\left[T\exp\left(ig\int_{t_1'}^{t_1} d\tau_1\sum_i A_{i,4}(\xi,\vec x_1,\tau_1)\right)\right]\cr
&&\times
\left[T\exp\left(ig\int_{ t_2'}^{ t_2} 
d\tau_2 \sum_i\bar A_{i,4}(\xi,\vec x_2,\tau_2)\right)\right], 
\label{W}
\end{eqnarray}
where the fields $ A_i^{(1)}$ and $ \bar A_i^{(2)}$ are the projections of the instanton fields to the 
lines $L_1$ and $L_2$ corresponding to the heavy quark $Q$ and  the heavy antiquark $\bar Q$, respectively.

Under the same argumentation as before (see Eq.~(\ref{w-1})), one may
 extend Pobyitca's Eq.~\cite{Pobylitsa:1989uq} and 
by neglecting ${\cal O}(N^2/V^2)$ terms get the solution 
of extended equation
\begin{eqnarray}
\label{W-11}
W^{-1}[\psi,\psi^\dagger]&=&\theta^{-1}\otimes\theta^{-1} 
-\frac{N}{2}\sum_\pm \frac{1}{\langle  V_{\pm}[\psi^{\dagger} ,\psi ]\rangle}\int d\zeta_\pm \prod_f V_{\pm,f}[\psi_f^{\dagger} ,\psi_f ]\cr
&&\times \left[\left(\theta^{-1}w^{(1)}_\pm\theta^{-1}\right)\otimes\left(\theta^{-1}\bar w^{(2)}_\pm\theta^{-1}\right)
-\theta^{-1}\otimes\theta^{-1}\right],
\end{eqnarray}
where $\otimes$ is tensor product. The second term in this 
Eq.~(\ref{W-11}) describes a heavy quark-antiquark $Q\bar Q$ pare 
interacting with 
$N_f$ light quarks (see Fig.~\ref{figQQq}).
\begin{figure}[!ht]\center
\tikz{\begin{feynman}
\vertex (i1){$Q$};
\vertex(q1) [below=0.7cm of i1]{$q_1$};
\vertex(m1) [below=0.4cm of q1];
\vertex(n1) [below=0.4cm of m1];
\vertex(o1) [below=0.4cm of n1]{$q_{N_f}$};
\vertex(j1) [below=0.7cm of o1]{$\bar Q$};
\vertex(c1) [right=2.2cm of i1]{$\rm $};
\vertex(c)[blob] [below=1.3cm of c1]{$\rm I$};
\vertex(i2) [right=2.2cm of c1]{$\bar Q$};
\vertex(o2) [below=0.7cm of i2]{$q_{N_f}$};
\vertex(n2) [below=0.4cm of o2];
\vertex(m2) [below=0.4cm of n2];
\vertex(q2) [below=0.4cm of m2]{$q_1$};
\vertex(j2) [below=0.7cm of q2]{$Q$};
\diagram*{
(i1) --[fermion,line width=1.5pt] (c) --[fermion,line width=1.5pt] (j2),
(j1) --[anti fermion,line width=1.5pt] (c) --[anti fermion,line width=1.5pt] (i2),
(q1) --[fermion] (c) --[fermion] (q2),
(o1) --[fermion] (c) --[fermion](o2);
};\end{feynman}}
\caption{Instanton generated heavy $Q\bar Q$ pare 
and $N_f$ light $q$ quarks interaction.}
		\label{figQQq}
	\end{figure}

In the ILM without account of light quarks
the application of  Eq.~(\ref{W-11}) provides the direct instanton 
contribution to the $Q\bar Q$ potential~\cite{Diakonov:1989un} 
as we discussed above
$$
V_{\rm dir}(r)=\frac{N}{2VN_c}\sum_{\pm} \int d^3z_\pm \tr_c
\left[1-P\exp\left( i\int_{L_1}
dt A_{\pm,4}\right)P\exp\left(- i\int_{L_2}
dt A_{\pm,4}\right)\right].
$$
At small distances ($x\rho^{-1}\ll 1$),  it can be evaluated
analytically and one has the following form potential
\begin{eqnarray}
V_{\rm dir}(r)&\simeq &\frac{4\pi\lambda}{N_c\rho}\,\left\{\frac{\pi^2}{3}\left[\frac{\pi}{16}-J_1(2\pi)\right]\frac{r^2}{\rho^2}\right.\cr
&-&\!\left.\pi\left[\frac{\pi^2(438+7\pi^2)}{30720}+\frac{J_2(2\pi)}{80}\right]\frac{r^4}{\rho^4}+\dots\!
\right\}\!,\qquad
\end{eqnarray}  
in terms of the Bessel functions $J_{n}$~\cite{Yakhshiev:2018juj,Musakhanov:2020hvk}. 
At the large values of the 
$Q\bar{Q}$ inter-distance ($r\gg \rho$),
the potential has the form 
\begin{equation}
V_{\rm dir}(r)\simeq 2\Delta M_Q^{\rm dir}-
\frac{2\pi^3\lambda}{N_c r}.
\end{equation}

An average size of charmonium  is comparable with 
the instanton size $r_c\sim\rho$ while for
botomonium the relation $r_b<\rho$ is hold.
Consequently, one may expect that $r^2$-approximation will work 
better in the botomonium case in comparison with the 
charmonium case. 
It is well know that $r^2$-approximation corresponds to 
the dipole approximation in the multi-pole expansion.

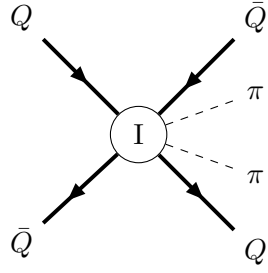
\begin{figure}[!ht]\center
\tikz{\begin{feynman}
\vertex (i1){$Q$};
\vertex (o1) [below=3cm of i1]{$\bar Q$};
\vertex(c)[blob] [below right=2.2cm of i1]{$\rm I$};
\vertex(i3) [below right=2.2cm of c]{$Q$};
\vertex(o2) [above right=2.2cm of c]{$\bar Q$};
\vertex(p1) [above=1cm of i3]{$\pi$};
\vertex(p2) [below=1cm of o2]{$\pi$};
\diagram*{
(i1) --[fermion,line width=1.5pt] (c) --[fermion,line width=1.5pt] (i3),
(o1) --[anti fermion,line width=1.5pt] (c) --[anti fermion,line width=1.5pt](o2),
(c)--[scalar](p1),
(c)--[scalar](p2);
};\end{feynman}}
\caption{Instanton generated emission of pions by heavy $Q\bar Q$ quark system.}
		\label{figQQpion}
	\end{figure}

The interaction term in the Eq.~(\ref{W-11}) has a part corresponding 
to the colorless state of light quarks. From this part we can 
calculate  the amplitude $A_{QQ\pi}$ of the process  
$(Q\bar Q)_{n'}\rightarrow (Q\bar Q)_n \pi\pi$ 
(see Fig.~\ref{figQQpion}) corresponding  to $N_f=2$ case 
\begin{eqnarray}
&&A_{QQ\pi}=F^2_{\pi Q}\int d^4z\,
\tr_f\partial_\mu U(z) \partial_\mu U^+(z)\exp(i(\vec p'-\vec p)\vec z)
\cr
&&\times\int d^3yd^3r\langle n|\vec r\rangle
\exp(-i\vec p\vec y)\frac{1}{N_c}\tr_c \left\{
1-P\exp\left(i\int_{-\infty}^{\infty}d\tau_1 A_{\pm,4}(\vec y+\vec r/2, \tau_1)\right)
\right.\cr
&&\left.\times\,P\exp\left(- i\int_{-\infty}^{\infty}d\tau_2 A_{\pm,4}(\vec y-\vec r/2, \tau_2)\right)\right\}
\exp(i\vec p^{\,\prime}\vec y)\langle\vec r|n'\rangle,
\label{AQQpions3}
\end{eqnarray}
where $\vec y=\vec x-\vec z$, the positions of $Q$ and $\bar Q$ are taken as $\vec x_1=\vec x+\vec r/2 $ and $\vec x_2=\vec x-\vec r/2 $. 
In Eq.\,(\ref{AQQpions3}),  $\exp(i\vec p^{\,\prime}\vec x)|n'\rangle$ and $\exp(i\vec p\vec x)|n\rangle$ 
are the initial and final states of $Q\bar Q$ system with the corresponding total 
momentums $\vec p^{\,\prime}=(\vec p^{\,\prime}_1+\vec p^{\,\prime}_2)$ and  $\vec p=(\vec p_1+\vec p_2)$, respectively.
They 
are solutions of the Schrodinger equation with the Hamiltonian 
\begin{eqnarray}
H_0=T+V,\qquad T=\frac{{\vec p_1}^{\,2}}{2m_Q}+\frac{{\vec p_2}^{\,2}}{2m_Q},
\label{HQQ}
\end{eqnarray}
where $V$  is ``$Q\bar Q$ potential in the istanton medium + phenomenological confining potential". 

The matrix element between the heavy quarkonium $Q\bar Q$ states 
in the amplitude~(\ref{AQQpions3}) has
the factor $ F(\vec r,\vec p^{\,\prime}-\vec p)$, which is given as
\begin{eqnarray}
\label{structure}
 F(\vec r,\vec p^{\,\prime}-\vec p)&=&\int d^3y\exp(i(\vec p^{\,\prime}
 -\vec p)\vec y)
\cr
&&\times\frac{1}{N_c}\tr_c\left\{1-P\exp\left( i\int_{-\infty}^{\infty}d\tau_1 A_{\pm,4}(\vec y+\vec r/2, \tau_1)\right)\right.\cr
&&\times\left.
P\exp\left(- i\int_{-\infty}^{\infty}d\tau_2 A_{\pm,4}(\vec y-\vec r/2, \tau_2)\right)\right\}.
\end{eqnarray}
From this equation we see, that 
$F(\vec r,\vec p^{\,\prime}-\vec p=0)={V}{N}^{-1}V_{\rm dir}(r)$.
For small $r\rho^{-1}$ we may apply an electric dipole approximation 
during the calculations of  
$ F(\vec r,\vec p^{\,\prime}-\vec p)$. As we already mentioned, the
 dipole approximation
may be well approximation in the botomonium case, while for the charmonium we expect sizable corrections from other terms of the expansion.

\subsection{
Standard approach and Phenomenology of the $(Q\bar Q)_{n'}\rightarrow (Q\bar Q)_n\,\, \pi\pi$ process}

According to Ref.\,\cite{Mannel1997} the 
phenomenological definition of the coupling for $(Q\bar Q)_{n'}\rightarrow (Q\bar Q)_n\, \pi\pi$  process
can be written in the form of effective lagrangian ${\cal L}$.
In the chiral $m_q\rightarrow 0$ and heavy quark mass $m_Q\rightarrow\infty$  limits 
it has the form
\begin{eqnarray}
{\cal L} = g A_\mu^{(v)} B^{(v)\mu *}
              \tr[(\partial_\nu U) (\partial^\nu U)^\dagger] +  h.c.
\end{eqnarray}
Here $A_\mu^{(v)}$ and $ B^{(v)\mu }$ are 
factors  corresponding to
$(Q\bar Q)_{n'}\sim (2S)$ and $ (Q\bar Q)_n\sim (1S)$ states. 
The experimental values of the couplings are given in Table\,\ref{Table4}.
\begin{table}[h]
\footnotesize
\caption{The values of coupling $g$ from the decay processes.}
\begin{center}
\begin{tabular}{@{}ccc}
\hline 
 & $\psi(2S) \rightarrow J/\psi\pi^+\pi^-$ & $\Upsilon(2S)
                 \rightarrow \Upsilon(1S)\pi^+\pi^-$\\
                 \hline
g   & $0.30 \pm 0.02$   & $0.25 \pm 0.02$   \\
\hline          
\end{tabular}
\end{center}
\label{Table4}
\end{table}
Our estimate  of $g_{J/\psi}$ in the framework of ILM is
\begin{eqnarray}
g_{J/\psi}=\frac{F^2_{\pi Q}}{F^2_\pi}1.345\frac{r_{J/\psi}^2}{\rho^2}\left( 1
-0.372\frac{r_{J/\psi}^2}{\rho^2}\right)
\label{g}
\end{eqnarray}
where the quantity outside of the bracket corresponds to the dipole 
approximation while the bracket  takes into account the next term 
corrections in the expansion over $(r/\rho)^2$.
Obviously, it is seen that $g_{\Upsilon} < g_{J/\psi}$.

\begin{figure}[h]
\begin{center}
\includegraphics[scale=0.5]{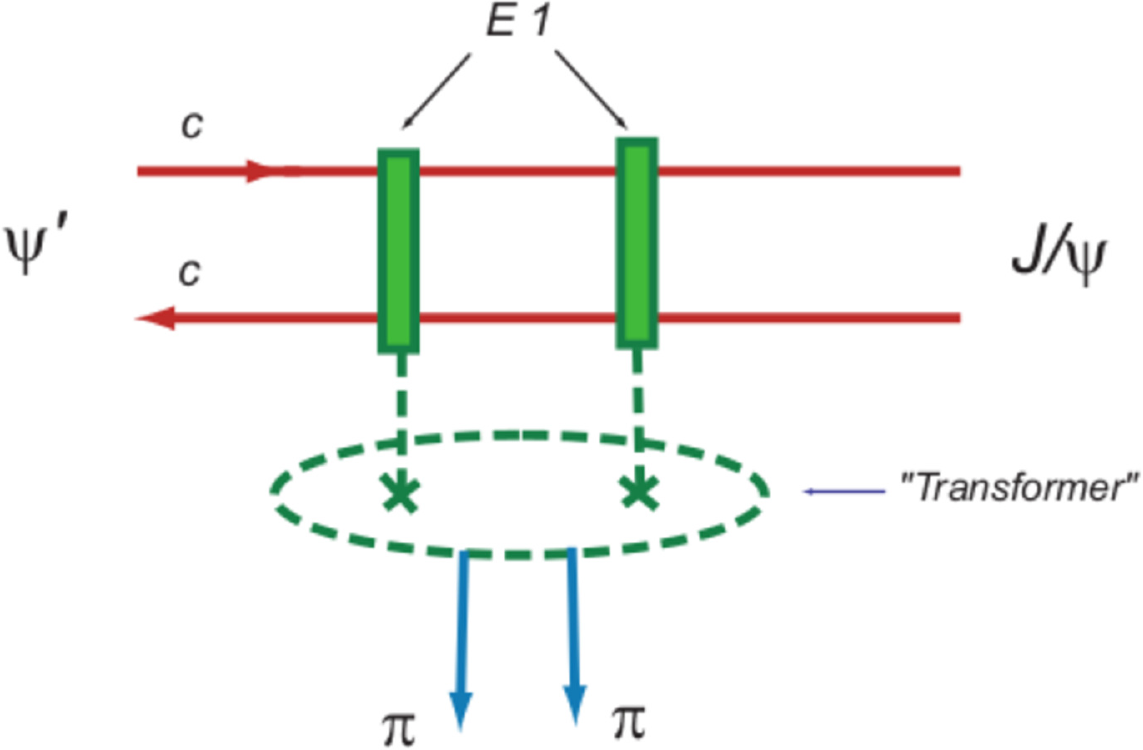}
\end{center}
\caption{Two pion production from the charmonia transitions.}
\label{Fig8}
\end{figure}

The standard approach to `the quarkonium -- light hadron transitions' 
assumes an applicability of the  multipole expansion, which means that 
the quarkonium sizes $r_{c,b}$ are much less than the typical size of
the nonperturbative vacuum gluon fluctuation $\lambda_g$ (see e.g.~\cite{Voloshin:2007dx}). 
According to this assumption the dipole approximation can be 
represented as shown in Fig.\,\ref{Fig8}.

However, in ILM $\lambda_g\approx\rho$. 
Using the instanton parameter $\rho = 1/3$~fm and  the charmonium size $r_{J/\psi}= 0.25$~fm (see Table\,\ref{Quarkoniumsizes}) in 
Eq.\,(\ref{g}) one can get the value $g_{J/\psi}=0.28(1-0.2)\simeq 0.22$. From here one can conclude, that the  correction to the dipole approximation in charmonium case is quite sizable, $\sim 20\%$.

\section{Summary}
\label{sec:SumOut}

Coming back to the spectrum of quarkonium one can note, 
that the exact comparisons of results with the 
experimental data can be done 
after inclusions of the spin-dependent parts in the 
potential. Nevertheless, one can make some qualitative 
predictions and corresponding conclusions. In Ref.\,\cite{Yakhshiev:2018juj} it was discussed
that the instanton background could explain 
the origin of potential model parameters.
For example, with the value of $\alpha_s=0.2068$
denoted as MWOI in Ref.\,\cite{Yakhshiev:2018juj} it was possible 
to fit the experimental data by using 
Cornell's type potential and concentrating on the 
first six low-lying S-wave states during the fitting process
(see Table\,II in Ref.\,\cite{Yakhshiev:2018juj} and 
explanations there). Although, the low-lying states 
were reproduced more or less
well the excited states came out lower estimated. Inclusion
of the direct instanton interactions improved much 
better the low-lying states but the excited states 
are came out overestimated (see columns M-I and M-IIb in
Table\,II of  Ref.\,\cite{Yakhshiev:2018juj}). 

The results of the work\,\cite{Musakhanov:2020hvk}
indicated that that the inclusion of screening 
effect from the instantons changes the situation
and  the screening 
effect softens the contributions to excited states from 
the instantons (see columns $V_{\rm scr}$
or ``$V_{\rm scr}$'' in Table\,\ref{Table3}).
As a result, the instanton effects 
from both $V_{\rm scr}$ and $V_{\rm dir}$ 
accumulated in such a way that the ground state
is changed in a different way while the excited 
states have more or less overall shift effect
(see columns $V_{\rm scr}+V_{\rm dir}$
or ``$V_{\rm scr}+V_{\rm dir}$''). This situation may 
be quite helpful in describing the experimental data
related to the charmonium states by using the potential 
approach in the framework of instanton liquid model. 
One can also note that, although the instanton effects are at
 the level of few percents they cannot be ignored. 
 In contrast they may be  important during the fine tuning processes of the whole spectrum. This studies 
 can be done after the inclusion relativistic corrections 
which takes into account the spin-spin, spin-orbit and tensor 
interactions. 

In this review we focused on the properties of 
gluons, light and heavy quarks in the instanton backgound. We 
emphasized that, although the instantons cannot explain 
a confinement mechanism they play a nontrivial role 
in both, nonperturbative and nonpeturbative regions.
 In the perturbative region the instantons will 
 produce the screening effect in the perturbative OGE 
 potential. At very short distances the OGE potential including the 
instanton effects have a Yukawa-type type nature due to the 
generation of the dynamical gluon mass. 
The direct instanton effects came out important 
in the nonperturbative region and produces an overall shift 
in the spectrum of quarkonia. One may conclude,
that the instanton effects in both, perturbative and 
nonperturbative, regions are important for understanding the heavy 
quark physics.

Summarizing the light-heavy systems nature in the instanton medium 
we note, that the instantons naturally generate also heavy-light quarks
interactions, which seem to be important for the heavy quark physics 
when the participation of light quarks becomes necessary. 
This is seen from our discussions, where we considered
exited heavy quarkonium decays with emission of pions. It is expected, 
that the applications of ILM to the  heavy-light quarks systems 
will affect their properties strongly.
\section*{Acknowledgments}
The work is supported by Uz grant OT-F2-10
(M.M.) and by the Basic Science Research Program 
through the National Research Foundation (NRF) of Korea funded 
by the Korean government (Ministry of Education, Science and
Technology, MEST), Grant Number~2020R1F1A1067876 (U.Y.).

\end{document}